# Title: Realization of a Dirac-vortex topological photonic crystal fiber

**Authors:** Quanhao Niu[1,2,3]†, Bei Yan[4]†*, Lei Shen[5]†, Hao Lin[6], Xi Zhang[1,2,3], Zhenyu Wan[1,2,3], Mutian Xu[1,2,3], Hui Zhang[5], Jie Luo[5], Lei Zhang[5], Perry Ping Shum[7]*, Zhen Gao[7]*, Jian Wang[1,2,3]*

**Affiliations:**

[1]Wuhan National Laboratory for Optoelectronics and School of Optical and Electronic Information, Huazhong University of Science and Technology, Wuhan 430074, Hubei, China

[2]Hubei Optical Fundamental Research Center, Wuhan 430074, China

[3]Optics Valley Laboratory, Wuhan 430074, Hubei, China

[4]Hubei Province Key Laboratory of Systems Science in Metallurgical Process, and College of Science, Wuhan University of Science and Technology, Wuhan 430081, China

[5]State Key Laboratory of Optical Fiber and Cable Manufacture Technology, Yangtze Optical Fiber and Cable Joint Stock Limited Company, Wuhan, Hubei 430074, China

[6]School of Physics and Optoelectronics, South China University of Technology, Guangzhou 510640, China

[7]State Key Laboratory of Optical Fiber and Cable Manufacture Technology, Department of Electronic and Electrical Engineering, Guangdong Key Laboratory of Integrated Optoelectronics Intellisense, Southern University of Science and Technology; Shenzhen 518055, China

†These authors contributed equally to this work

*Corresponding author: jwang@hust.edu.cn, gaoz@sustech.edu.cn, shum@ieee.org, yanbei@wust.edu.cn

**Abstract:** Photonic crystal fibers (PCFs) that trap and guide light using photonic bandgaps have revolutionized modern optics with enormous scientific innovations and technological applications spanning many disciplines. Recently, inspired by the discovery of topological phases of matter, Dirac-vortex topological PCFs have been theoretically proposed with intriguing topological properties and unprecedented opportunities in optical fiber communications. However, due to the substantial challenges of fabrication and characterization, experimental demonstration of Dirac-vortex topological PCFs has thus far remained elusive. Here, we report the experimental realization of a Dirac-vortex topological PCF using the standard stack-and-draw fabrication process with silica glass capillaries. Moreover, we experimentally observe that Dirac-vortex single-polarization single-mode bounds to and propagates along the fiber core in the full communication window (1260-1675nm). Our study pushes the research frontier of PCFs and provides a new avenue to enhance their performance and functionality further.

**Main Text:** Since its discovery in 1991, photonic crystal fibers (PCFs) [1,2] that use photonic bandgaps to confine and guide light have attracted great attention across many disciplines due to their outstanding ability to overcome the limitations of conventional optical fibers, such as bending-insensitive low-loss guidance of light, design flexibility of fiber cross-sections, and high-power light delivery. Recently, inspired by the advancements in topological photonics [3-9], the concept of topological PCFs with robust topological protection has been theoretically proposed [10-17] based on different topological physical mechanisms and experimentally demonstrated [18,19] in a multicore PCF whose cores are arranged in a Su-Schrieffer-Heeger chain with topological end states, significantly improving the performance and functionality of conventional PCFs and exhibiting promising applications in next-generation robust quantum network and optical communications. In particular, Dirac-vortex topological PCFs [7], originating from Dirac-vortex states in a Kekulé-distorted honeycomb lattice [20-30], have been theoretically proposed with ultra-broadband single-polarization single-mode light transport or arbitrary degenerate fiber modes. However, to date, Dirac-vortex topological PCFs have remained out of experimental reach.

In this work, we fabricated and experimentally demonstrated, for the first time to our best knowledge, the Dirac-vortex topological PCF using the standard stack-and-draw process with silica glass tubes. In particular, we directly observed that the Dirac-vortex mode bounds to and propagates along the fiber core, with its unique intensity profile and polarization distribution consistent with the theoretical results in a wide wavelength range (1260-1675nm). Moreover, we experimentally explored the influence of different incident light beams on the fiber output, exhibiting excellent and robust single-polarization and single-mode property of the Dirac-vortex topological PCF.

**Results**

We begin with a common silica PCF with perfect triangular lattice of air holes and introduce Kekulé distortion to the thickness of the silica struts in the cross section to construct a Dirac-vortex topological PCF, as illustrated in Fig. 1A. Fig. 1B shows a unit cell of the cross section of the perfect triangular PCF with a lattice constant of $\sqrt{3}a$ ($a$ = 3μm) and a silica strut width of $t_0 = 0.16a$. The Kekulé distortion is introduced by changing the strut width with Dirac mass $m = m_0 e^{i\varphi}$, where $m_0$ is the amplitude of the Dirac mass, $\varphi = \omega\theta(r)$ is the vortex phase angle, $\theta(r)$ is the space phase angle, and $\omega$ is the vortex winding number. After Kekulé distortion, the strut width becomes $t_1 = t_0 + m_0\cos(\varphi)$ (red color), $t_2 = t_0 + m_0\cos(\varphi+2\pi/3)$ (blue color), and $t_3 = t_0 + m_0\cos(\varphi+4\pi/3)$ (green color), where $t_0 = 0.16a$ and winding number $\omega = 1$, as shown in Fig. 1C. When the Dirac mass $m_0 = 0$ (no Kekulé distortion), the bulk band structure of the cross section of the perfect triangular PCF (grey lines) exhibits a double Dirac point with fourfold degeneracy at the Γ point with nonzero wave vector $k_z$. When $m_0 \neq 0$ and $\varphi = \pi/3$ (with Kekulé distortion), the degeneracy will be lifted (red lines) and open a complete topological photonic bandgap at $k_z = 4\pi/a$ (see the band inversion and Wilson loop in the Supplementary Materials), as shown in Fig. 1D. More significantly, the topological photonic bandgap persists for arbitrary phase of $\varphi$ if $m_0 \neq 0$, as shown in Fig. 1E, in which the color represents the phase $\varphi$ varying from 0 to $2\pi$. We then arrange a series of Kekulé-distorted triangular photonic crystals with six different modulation phases $\varphi$ angularly around the fiber core to form a Dirac-vortex topological PCF, as shown in Fig. 1F. As a result, a single Dirac-vortex topological PCF mode appears in the middle of the bandgap, being tightly localized, and propagates along the fiber core with nonzero wave vector $k_z$.

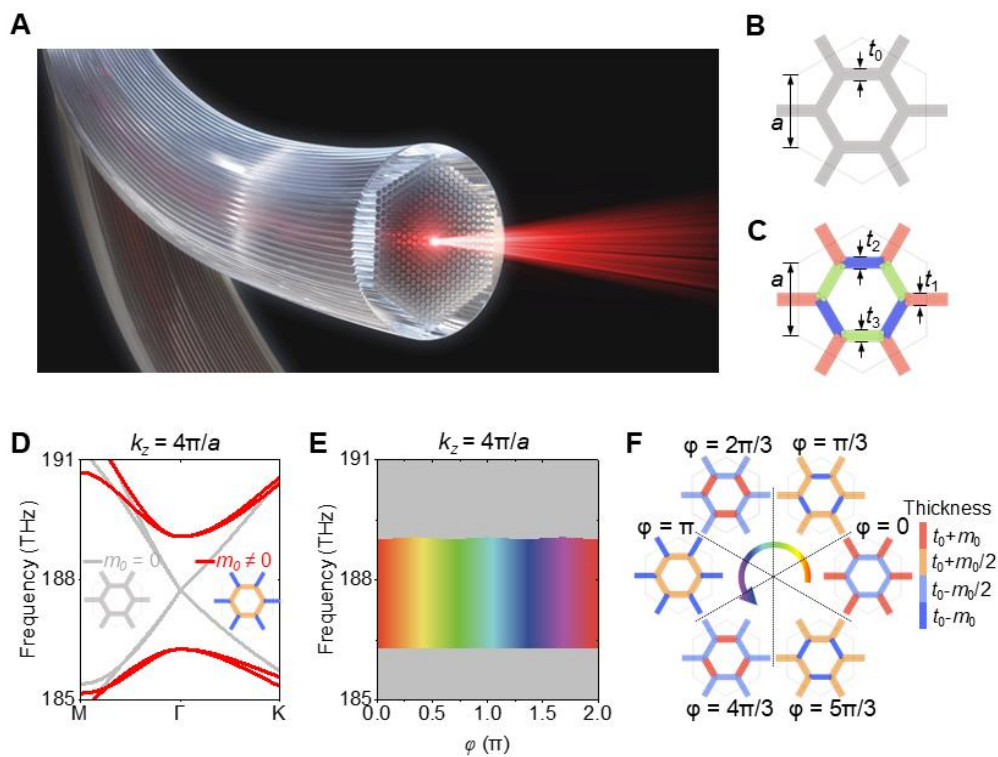

**Fig. 1 | Designing a Dirac-vortex topological PCF by introducing Kekulé distortion in a triangular PCF. A** Schematic of the Dirac-vortex topological PCF. **B, C** The unit cell of the cross-section of a triangular PCF without (**B**) and with (**C**) Kekulé distortion. The Kekulé distortion is introduced by changing the strut width with $m_0$ and phase $\varphi$, where $t_1 = t_0+m_0\cos(\varphi)$, $t_2 = t_0+m_0\cos(\varphi+2\pi/3)$, $t_3 = t_0+m_0\cos(\varphi+4\pi/3)$, $t_0 = 0.16a$, and $a = 3\mu m$. **D** Simulated bulk band structures of the cross section of the triangular PCF without (grey lines) and with (red lines) Kekulé distortion at $k_z = 4\pi/a$, respectively. **E** Vortex bandgaps with fixed Dirac mass $m_0 = 0.024a$ and different phases $\varphi$ at $k_z = 4\pi/a$. The color represents the phase $\varphi$ varying from 0 to $2\pi$. **F** Schematic of the Dirac-vortex topological PCF consisting of six aperiodic Kekulé-distorted triangular photonic crystals with a winding number of +1. The color of the struts represents their widths.

Fig. 2A shows the cross section of the Dirac-vortex topological PCF with space-dependent six different modulation phases $\varphi$ and fixed $m_0 = 0.024a$ angularly around the fiber core. From both the fiber dispersion with absolute frequency (Fig. 2B) and relative frequency (Fig. 2C), we can observe a single Dirac-vortex topological PCF mode (red line) in a broad frequency range of 164-317THz (946-1829nm) located in the topological bandgap induced by the Kekulé distortion. Besides the nontrivial Dirac-vortex topological PCF mode, there also exist trivial index-guided modes (blue line) in the Dirac-vortex fiber whose frequency is the lowest for a common wave vector, usually appearing in pairs with different polarizations and confined in wherever there is a local maximum of the strut thickness (equivalent to a high effective refractive index). Here we only focus on the Dirac-vortex fiber modes inside the topological bandgap, as shown in Fig. 2D, whose field intensity (red color) and electric vector (black arrows) distribution are tightly localized around the fiber core with a $C_{3v}$ symmetry, similar to an azimuthally polarized beam (APB). To facilitate the fabrication of the Dirac-vortex topological PCF, we round off the six sharp corners of the hexagon with a radius of $r$, as shown in the inset of Fig. 2E. As we increase the radius $r$, the topological bandgap (grey region) gradually decreases but always exist as long as the hexagonal air holes do not transform into circular air holes. Fig. 3F shows the eigenmode spectrum of a Dirac-vortex topological PCF with a radius $r = 0.3a$, from which we can see a Dirac-vortex topological PCF mode (red dot) exists in the topological bandgap, whose field intensity is almost the same as that in Fig. 2D.

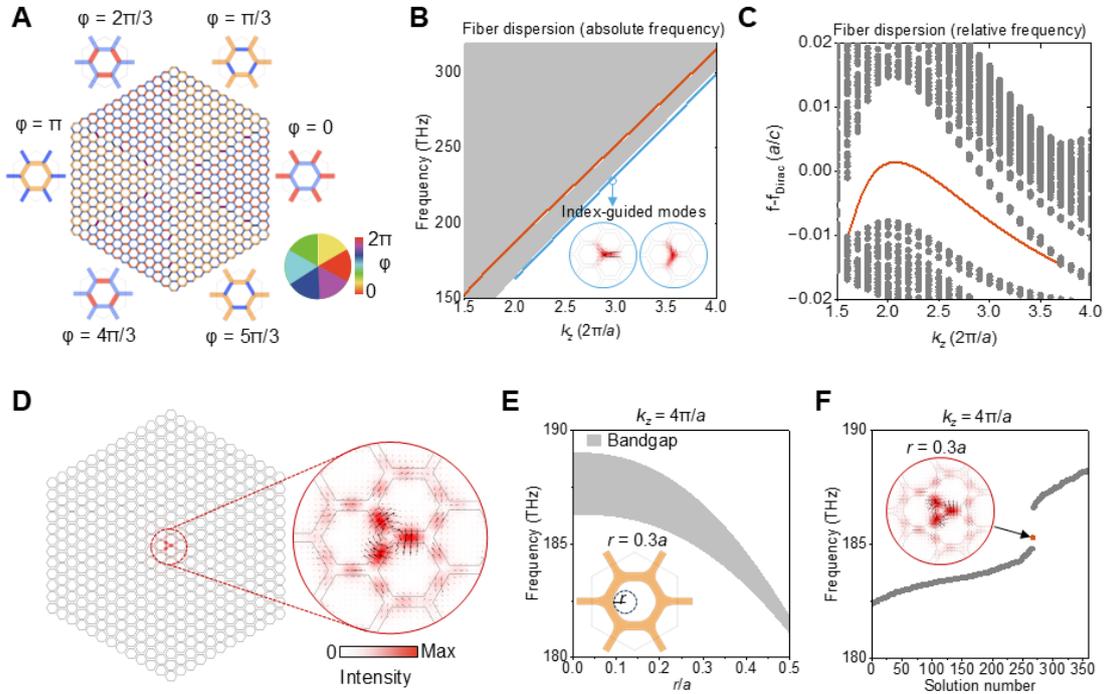

**Fig. 2 | Dirac-vortex topological PCF**. **A** Schematic of the Dirac-vortex topological PCF cross-section. The inset color bar represents the phase distribution of the Kekulé modulation. **B, C** Simulated fiber dispersion with absolute (**B**) and relative (**C**) frequencies, and the red (blue) line represents the Dirac-vortex fiber (index-guided) modes. The intensity (red color) and electric vector (black arrows) distributions of the index-guided modes are shown in the inset of **B**. **D** Simulated intensity (red color) and electric vector (black arrows) distributions of the topological Dirac-vortex fiber modes at $k_z = 4\pi/a$. **E** Topological Dirac-vortex bandgap as a function of the radius $r$ of the rounded corners at $k_z = 4\pi/a$. Inset: a unit cell with $r = 0.3a$ and $t_0 = 0.16a$. **F** Simulated eigenstate spectrum of the Dirac-vortex topological PCF with $r = 0.3a$. Inset: simulated intensity (red color) and electric vector (black arrows) distributions of the topological Dirac-vortex fiber modes with $r = 0.3a$ at $k_z = 4\pi/a$.

Next, we fabricate the Dirac-vortex topological PCF using the standard stack-and-draw technique (see detailed fabrication process in Supplementary Material). We first stack four kinds of silica glass capillaries with the same outer diameter $d$ and different inner diameters to form the fiber preform. The combination of silica glass capillaries with inner diameters $d_1$ and $d_2$ ($d_3$ and $d_4$) can be used to construct the unit cell with modulation phase $\varphi = 0, 2\pi/3, 4\pi/3$ ($\varphi = \pi/3, \pi, 5\pi/3$), respectively, which consist of struts with thickness $t_0 \pm m_0/2$ ($t_0 \pm m_0$), as illustrated in Fig. 1F (see detailed correspondence between the combination of four different silica capillaries and the varying strut widths in Supplementary Material). Then we draw the stacked silica glass capillary structure using a fiber drawing tower to produce a uniformly sized Dirac-vortex topological PCF, as shown in Fig.3A. The difference between the actual and theoretical geometrical parameters is less than 0.1μm, indicating the stack-and-draw technique is stable and suitable to fabricate the Dirac-vortex topological PCF.

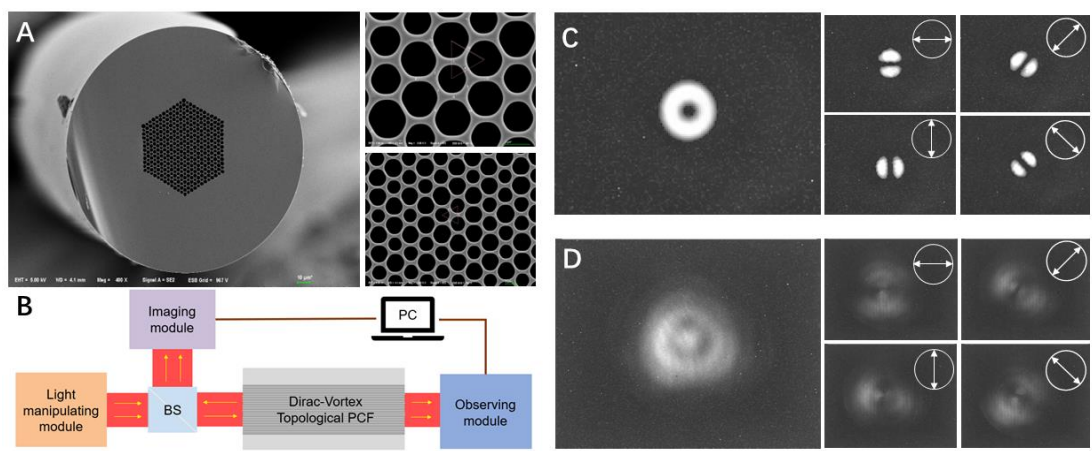

**Fig. 3 | Fabrication and characterization of Dirac-vortex topological PCF**. **A** Photograph and magnified image of the cross-section of the Dirac-vortex topological PCF. **B** Schematic diagram of the measurement of Dirac-vortex topological PCF. **C** The input APB and different polarization states detected in the light manipulating module. **D** The output Dirac-vortex topological PCF beam and different polarization states detected in the observing module.

We then experimentally characterize the fabricated Dirac-vortex topological PCF. The simplified diagram of experimental setup is illustrated in Fig. 3B (see detailed experimental configuration in Supplementary Material). The light manipulating module consists of a Sagnac interferometer with a spatial light modulator (SLM) to generate arbitrarily shaped scalar and vectorial modes as the incident beam. The imaging module, composed of an illuminating source and an imaging system, is placed along the optical field coupling path to monitor the fiber input facet. The observing module is adopted to map the polarization-intensity patterns of the output beam of the Dirac-vortex topological PCF. To match the topological mode supported by the Dirac-vortex fiber, we first use an APB beam created by the light manipulating module as the input beam at 1550nm, as shown in the left panel of Fig. 3C, which is carefully aligned to the fiber core with beam size reduction by a long focal lens combined with an objective lens to match the effective mode field area of the Dirac-vortex topological PCF. By rotating the polarizer (white arrows), the intensity patterns of the APB turn into two symmetric petals whose orientation is altered with different polarizations, as shown in the right panel of Fig. 3C. The measured intensity pattern of the output Dirac-vortex topological PCF mode is shown in the left panel of Fig. 3D, matching well with the simulation results shown in Fig. 2D and Fig. 2F. When the output Dirac-vortex topological PCF modes are passed through different oriented polarizers (white arrows), the intensity patterns of the Dirac-vortex fiber modes exhibit similar distributions with those of the APB, verifying the unique vectorial nature of the Dirac-vortex topological PCF modes.

To investigate the broadband single-polarization single-mode property of the Dirac-vortex topological PCF, three different input beams, including linearly polarized (LP) Gaussian beam, circularly polarized (CP) Gaussian beam, and the first-order orbital angular momentum ($OAM_{+1}$) mode, are coupled into the Dirac-vortex topological PCF, respectively, as shown in the left panel of Fig. 4A. The corresponding measured output intensity patterns without and with different oriented polarizer (white arrows) are presented in the right panel of Fig. 4A. It can be seen that the output fields remain the Dirac-vortex fiber mode even under different incident beams. Fig. 4B shows the measured relative output power at various wavelengths under different incident beams. For each incident beam, the power variation is within 1 dB, suggesting that the Dirac-vortex topological PCF exhibits a broadband and stable performance across the full communication window (1260-1675 nm). Moreover, the output powers of the Dirac-vortex fiber under APB (blue line) and $OAM_{+1}$ (red line) beams are much higher than those under CP (brown line) and LP (green line) Gaussian beams, indicating the more closely the input beam aligns with the topological Dirac-vortex fiber mode, the higher the output power. Remarkably, although the APB and $OAM_{+1}$ beams both have similar doughnut intensity patterns, the output power of the Dirac-vortex fiber under APB (blue line) is higher than that under $OAM_{+1}$ (red line). This can be explained with the fact that the APB beam with azimuthal polarization distribution is closer to the topological Dirac-vortex fiber mode. To further demonstrate the broadband single-polarization single-mode characteristic of the Dirac-vortex topological PCF, we also plot the measured output optical fields without and with different oriented polarizer (white arrows) at different wavelengths, as shown in Fig. 4C. We observe that the output topological fiber modes keep almost the same across the full communication window (1260-1675nm), verifying that the Dirac-vortex fiber maintains its topological properties over a broad wavelength range. It is worth noting that the measured 415 nm wavelength range (1260-1675 nm) is mainly based on the current available lab conditions. From the trend of the curves in Fig. 4B, there should be a larger bandwidth of the Dirac-vortex topological PCF, as predicted by theory, indicating its superior performance in an ultra-broad wavelength range.

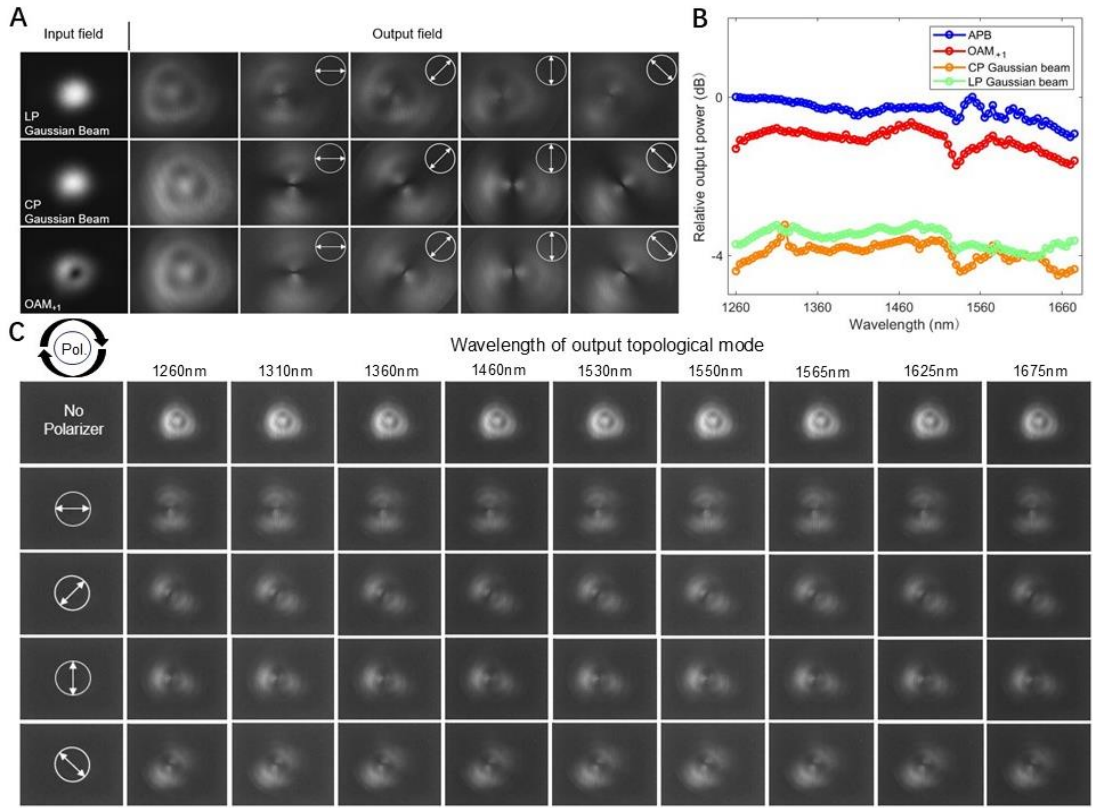

**Fig. 4 | Measured output field intensity distributions of the Dirac-vortex topological PCF under different incident beams and wavelengths. A** Measured input (left panel) and output (right panel) field intensity distributions of the Dirac-vortex topological PCF under incident LP Gaussian beam (top row), CP Gaussian beam (middle row), and the first-order $OAM_{+1}$ beam (bottom row). The white arrows represent different oriented polarizers. **B** Measured relative output power of the Dirac-vortex topological PCF under incident APB beam (blue line), $OAM_{+1}$ beam (red line), CP Gaussian beam (brown line), and LP Gaussian beam (green line) across the full 415 nm communication window (1260-1675nm). **C** Measured output field intensity distributions of the Dirac-vortex topological PCF without (top row) and with (lower rows) different oriented polarizers (white arrows) at 1260, 1310, 1360, 1460, 1530, 1550, 1565, 1625, and 1675nm, respectively.

**Discussion**

In summary, we have fabricated a Dirac-vortex topological PCF that supports broadband single-polarization single-mode light transport. We show that the Dirac-vortex fiber modes are robust against deformation induced in the stack-and-draw fabrication process and different incident beams, making them well-suited for robust optical signal transport. It would also be interesting to experimentally explore Dirac-vortex topological PCF with arbitrary near-degenerate modes by simply increasing the winding number of the vortex and tunable effective mode area by changing the vortex size, as well as its topological protection against fabrication imperfections and bending losses. Our work may enable new fiber applications with robust topological protections, such as high-power topological fiber lasers, quantum networks with protected entangled states of light, and stable optical fiber communications. We envision that enormous new topological phenomena and applications will emerge by introducing non-Hermitian, non-linear, non-abelian, and non-crystalline effects into topological PCFs.

In preparing our manuscript, we noticed two related works demonstrating topological PCFs based on helically tested twisted fiber [*31*] and disclination defect [*32*].